\documentclass{article}

\usepackage[top=1.0in, bottom=1.25in, left=0.875in, right=0.875in]{geometry}

\usepackage[cmex10]{amsmath}

\usepackage{amssymb}
\usepackage{graphicx}
\usepackage{float}
\usepackage{tabularx}
\usepackage{multirow}
\usepackage{authblk}  
\usepackage{subcaption}
\usepackage{rotating}  

\let\oldtabular\tabular
\renewcommand{\tabular}{\footnotesize\oldtabular}

\begin{document}
\title{Non-contact transmittance photoplethysmographic imaging (PPGI) for long-distance cardiovascular monitoring}
\date{}

\author[1,*]{Robert Amelard}
\author[1]{Christian Scharfenberger}
\author[1]{Farnoud Kazemzadeh}
\author[2]{Kaylen J Pfisterer}
\author[3]{Bill S Lin}
\author[1]{Alexander Wong}
\author[1]{David A Clausi}
\affil[1]{University of Waterloo, Department of Systems Design Engineering, Waterloo, Canada, N2L3G1}
\affil[2]{University of Waterloo, Department of Kinesiology, Waterloo, Canada, N2L3G1}
\affil[3]{University of Waterloo, Department of Mechanical and Mechatronics Engineering, Waterloo, Canada, N2L3G1}

\maketitle

%
%
\begin{abstract}
Photoplethysmography (PPG) devices are widely used for monitoring cardiovascular function. However, these devices require skin contact, which restrict their use to at-rest short-term monitoring using single-point measurements. Photoplethysmographic imaging (PPGI) has been recently proposed as a non-contact monitoring alternative by measuring blood pulse signals across a spatial region of interest. Existing systems operate in reflectance mode, of which many are limited to short-distance monitoring and are prone to temporal changes in ambient illumination. This paper is the first study to investigate the feasibility of long-distance non-contact cardiovascular monitoring at the supermeter level using transmittance PPGI. For this purpose, a novel PPGI system was designed at the hardware and software level using ambient correction via temporally coded illumination (TCI) and signal processing for PPGI signal extraction. Experimental results show that the processing steps yield a substantially more pulsatile PPGI signal than the raw acquired signal, resulting in statistically significant increases in correlation to ground-truth PPG in both short- ($p \in [<0.0001, 0.040]$) and long-distance ($p \in [<0.0001, 0.056]$) monitoring. The results support the hypothesis that long-distance heart rate monitoring is feasible using transmittance PPGI, allowing for new possibilities of monitoring cardiovascular function in a non-contact manner.
\end{abstract}

\flushbottom
%
%

%
%
\section*{Introduction}
Photoplethysmography (PPG) is a non-invasive light-based method that has been used since the 1930s for monitoring cardiovascular activity~\cite{hertzman1937,allen2007,shelley2007}. These devices have been used to assess cardiovascular factors such as blood oxygen saturation, heart rate, autonomic function, and peripheral vascular disease~\cite{allen2007}. Conventional contact PPG devices are fastened to the skin using a spring-loaded clamp onto peripheral sites including the finger, ear, and toe~\cite{allen2007}. Standard devices are comprised of either a single or multiple light-emitting diodes (LED) and a photodetector, and monitor cardiovascular activity by evaluating the change in light intensity due to fluctuations in blood volume. However, contact PPG devices are limited to monitoring a single individual per device, resulting in a linear increase in cost per individual monitored, and limiting the number of concurrent measurements. Furthermore, these devices produce single-point blood pulse measurements, which does not provide an indication for blood perfusion patterns which may be important to tissue viability~\cite{levy2008}. Finally, their contact nature makes them unsuitable for long-term continuous monitoring due to comfortability and movement artifacts~\cite{allen2007}.

To address these issues, non-contact photoplethysmographic imaging (PPGI) systems have been proposed. The goal of a PPGI system is to extract the blood pulse signal without tissue contact. Although existing designs differ, they are primarily comprised of the same primary components as contact PPG: a light source (LED) and a light detector (camera). Such systems rely either on active~\cite{wieringa2005,humphreys2007,zheng2008,cennini2010,kamshilin2011,sun2011} or ambient tissue illumination~\cite{verkruysse2008,poh2010,sun2012,kong2013,xu2014}. Both active and ambient PPGI systems are sensitive to temporal changes in uncontrolled ambient illumination. The severity of this effect on the system's ability to extract the subtle blood pulse signal is apparent with studies employing data collection in dark room settings~\cite{wieringa2005,zheng2008,cennini2010,sun2011}. Normalizing ambient illumination changes using software has been proposed~\cite{xu2014}, however this technique relies on a spectral estimation of the ambient illumination, which may fail in difficult lighting conditions.

A potential, yet largely uninvestigated, advantage of PPGI is the ability to perform long-distance monitoring. Long-distance monitoring was defined as monitoring at the supermeter level. Long-distance monitoring is difficult or infeasible with contact PPG devices, since they are either attached via a cable to a monitor, must store the data on the device, or must transmit the data wirelessly, resulting in the need for a specialized network infrastructure. Still, one device can only monitor one individual. In contrast, PPGI systems may acquire long-distance measurements of multiple individuals, provided that the individuals are in the camera's field of view. However, long-distance monitoring becomes challenging with existing systems. Many existing PPGI systems operate in reflectance mode, where the camera and illumination are positioned on the same side of the tissue under investigation. In addition to the weakened reflectance signal by the highly scattering nature of skin~\cite{anderson1981,gibson2005}, reflectance PPGI systems using divergent light sources are sensitive to a strong illumination intensity decrease. These factors render long-distance monitoring a challenging problem. These constraints may be alleviated using a transmittance PPGI system, where the LED is positioned close to a thin anatomical location (e.g., finger, toe). To the best of the authors' knowledge, this paper is the first to investigate long-distance transmittance PPGI monitoring.

\begin{figure}[H]
\centering
\includegraphics[width=0.85\textwidth]{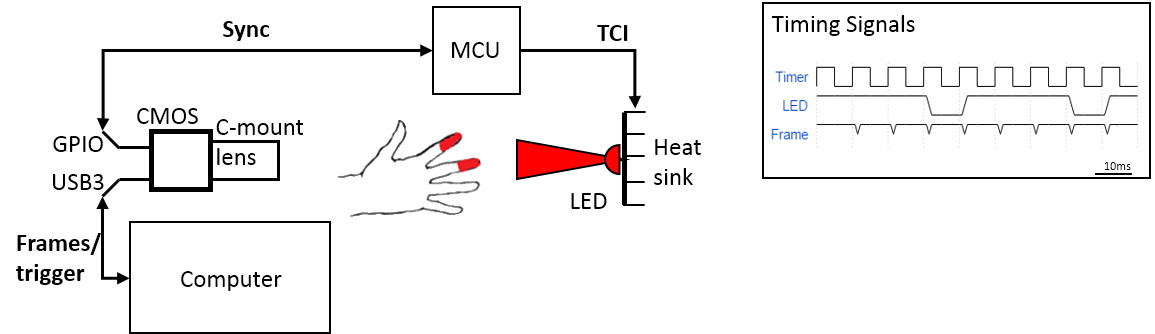}
\caption{The proposed non-contact PPGI system comprises an LED (655~nm), a CMOS camera (100~fps), and a microcontroller (MCU) for synchronization and TCI implementation. The MCU timer dictates the TCI code, triggering the camera to capture frames, which are transmitted and processed on the computer.}
\label{fig:system_diagram}
\end{figure}

\begin{figure}[H]
\centering
\includegraphics[width=\textwidth]{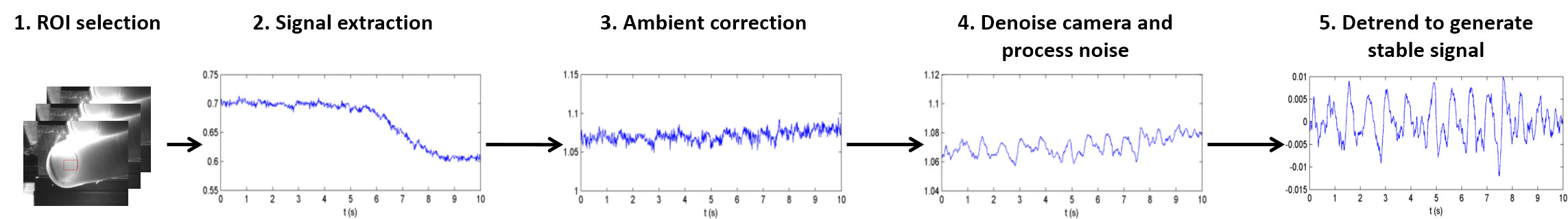}
\caption{Overview of the processing steps for the proposed PPGI system. Upon capturing frames, ambient correction using TCI is performed, which removes temporal changes in ambient illumination from the frames. This is followed by denoising to remove camera sensor noise and process noise from the light-tissue interaction, and detrending to provide a stable PPGI signal. The resulting signal is an extracted stable PPGI signal.}
\label{fig:system_flow}
\end{figure}

This paper presents a pilot study to assess the feasibility of long-distance cardiovascular monitoring using transmittance PPGI. For this purpose, a novel non-contact transmittance PPGI system is proposed which is able to monitor cardiovascular activity remotely by correcting for ambient lighting fluctuations and extracting the subtle blood pulse signal using signal processing tools. This system comprises a 100~fps camera, a high-powered LED (655~nm), a microcontroller to synchronize frame captures and illumination, and image and signal processing software on a computer. Figure~\ref{fig:system_diagram} shows a pictoral representation of this system, and Figure~\ref{fig:system_flow} shows the processing steps. Temporally coded illumination (TCI) is introduced to remove ambient lighting artifacts at the acquisition level, thus correcting the data prior to further processing, similar to that introduced in our previous work~\cite{amelard2015}. The proposed PPGI system used for this study extends substantially beyond our previously reported work by incorporating a comprehensive set of signal processing steps, a larger testing sample size, and more rigorous statistical and qualitative analysis. The signal processing is particularly important for this study, as long-distance cardiovascular monitoring scenarios result in acquired signals with low signal-to-noise ratios (SNR).


\section*{Results}
\label{sec:results}
\subsection*{Experimental Setup}
For this pilot study, five of the authors' data (age $28.2 \pm 5.3$) were collected using the proposed system. A 3:1 TCI code (alternating 3~frames with LED on, followed by 1~frame with LED off) was used with 100~fps frame rate and 8~ms exposure time. The aperture was manually adjusted to limit pixel saturation. The EasyPulse PPG device~\cite{easypulse} was used to collect the ground-truth PPG signal for validation.

\begin{figure}[H]
\centering
\includegraphics[width=0.6\textwidth]{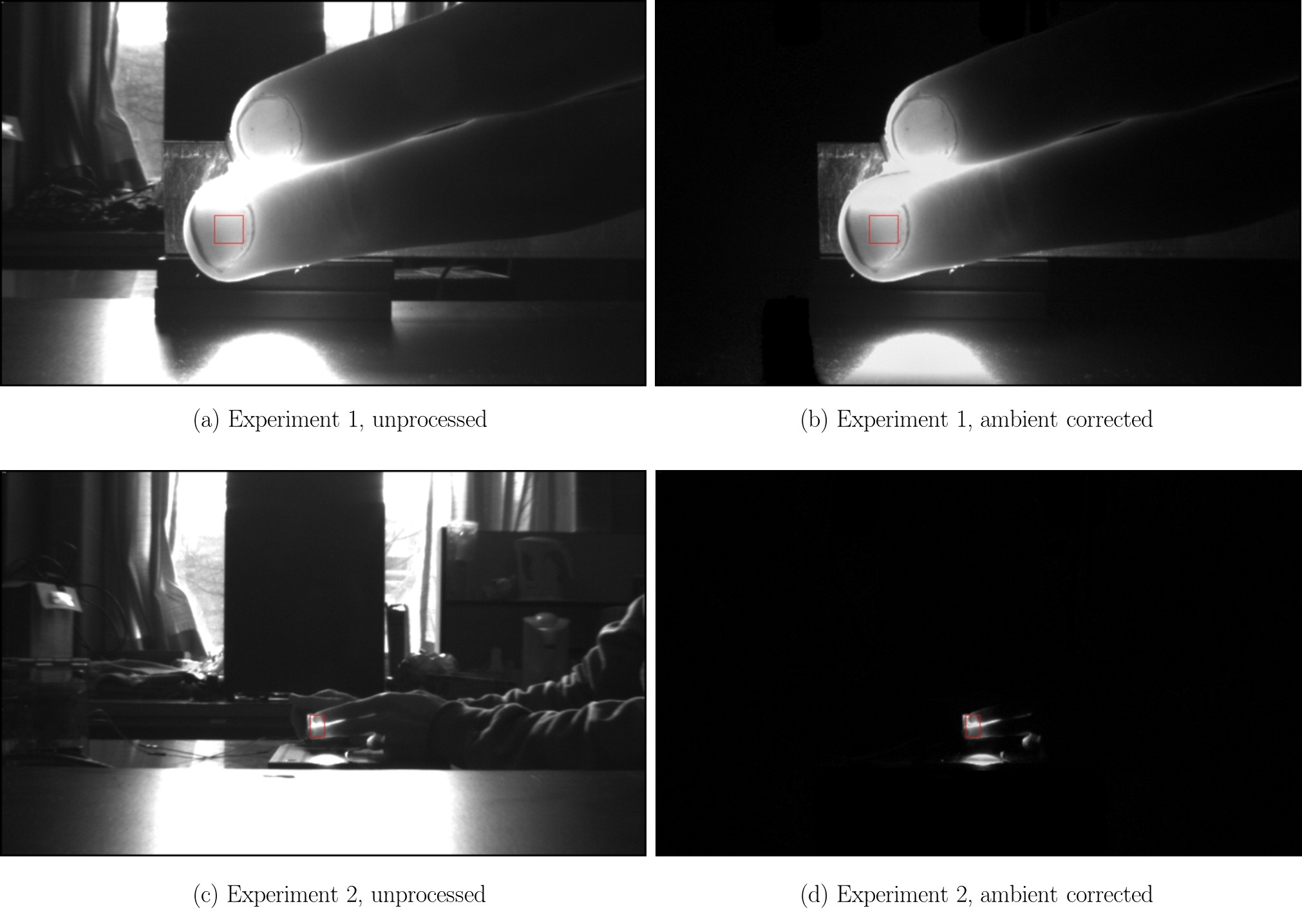}
\caption{Example images for Experiment 1 (short-distance monitoring) and Experiment~2 (long-distance monitoring). The unprocessed frames (first column) contain uncontrolled ambient illumination (windows, overhead lights, etc.) as well as controlled active LED illumination near the fingers. Ambient correction using TCI (second column) removes the contribution of ambient illumination to the scene, yielding transmittance due solely to active LED illumination of which the spectral and power characteristics are known. See Media~1 for video results of ambient correction with varying illumination.}
\label{fig:results_imgs}
\end{figure}

Two experiments were performed to assess the feasibility of long-distance monitoring. In Experiment~1, the camera and LED were separated by 20~cm, serving as short-distance base case validation. The participants were asked to position their fingers between the LED and camera so that their fingers covered the beam of the LED. In Experiment~2, the camera and LED were separated by 1.5~m (``long-distance''), and the participants were asked to position their fingers at approximately 10~cm from the LED. Figure~\ref{fig:results_imgs} shows example images of both experiments before and after ambient correction. For each experiment, a 10~s window was chosen that yielded a clean ground-truth PPG signal for validation. The normalized power spectral density (PSD) was computed for spectral analysis to demonstrate each signal's dominant frequency components. In an ideal signal, the fundamental heart rate should be the dominant frequency. Furthermore, to assess temporal signal fidelity, the Pearson's linear correlation coefficient $\rho$ was computed between between PPGI and PPG signals. This metric is offset- and scale-invariant, suitable for the problem of comparing unit-less PPG signals:
\begin{equation}
  \rho(X,Y)=\frac{\sigma_{XY}}{\sigma_X\sigma_Y} \in [-1,1]
\end{equation}
where $X,Y$ are the finger PPG and PPGI signal, $\sigma_{XY}$ is the covariance between the two signals, and $\sigma_X,\sigma_Y$ are the standard deviations of the same variables. This generates a value $\rho \in [-1,1]$, where $1$ is perfect linear correlation, $-1$ is a perfect negative linear correlation, and $0$ is uncorrelated. A perfectly reconstructed PPGI signal would yield $\rho=1$, indicating that the PPGI signal is a linear multiple of the finger PPG signal.

The PPGI signals were statistically analysed using a novel physiologically-motivated $t$-test to quantify the degree to which the proposed system processing improved the signal. For each participant, each heart beat instance (i.e., diastole-to-diastole) was semi-automatically labeled using a custom gradient descent approach. $\rho$ was computed for each such heart beat instance, yielding $n$ correlation values. A two-tailed paired-sample $t$-test ($\alpha=0.05$) was conducted using the correlation values for each heart beat instance, yielding a $p$-value indicating the probability of a zero-mean difference between signal correlation.

\subsection*{Experiment 1: Short-Distance}
\begin{sidewaysfigure}
\centering
\includegraphics[width=0.95\textwidth]{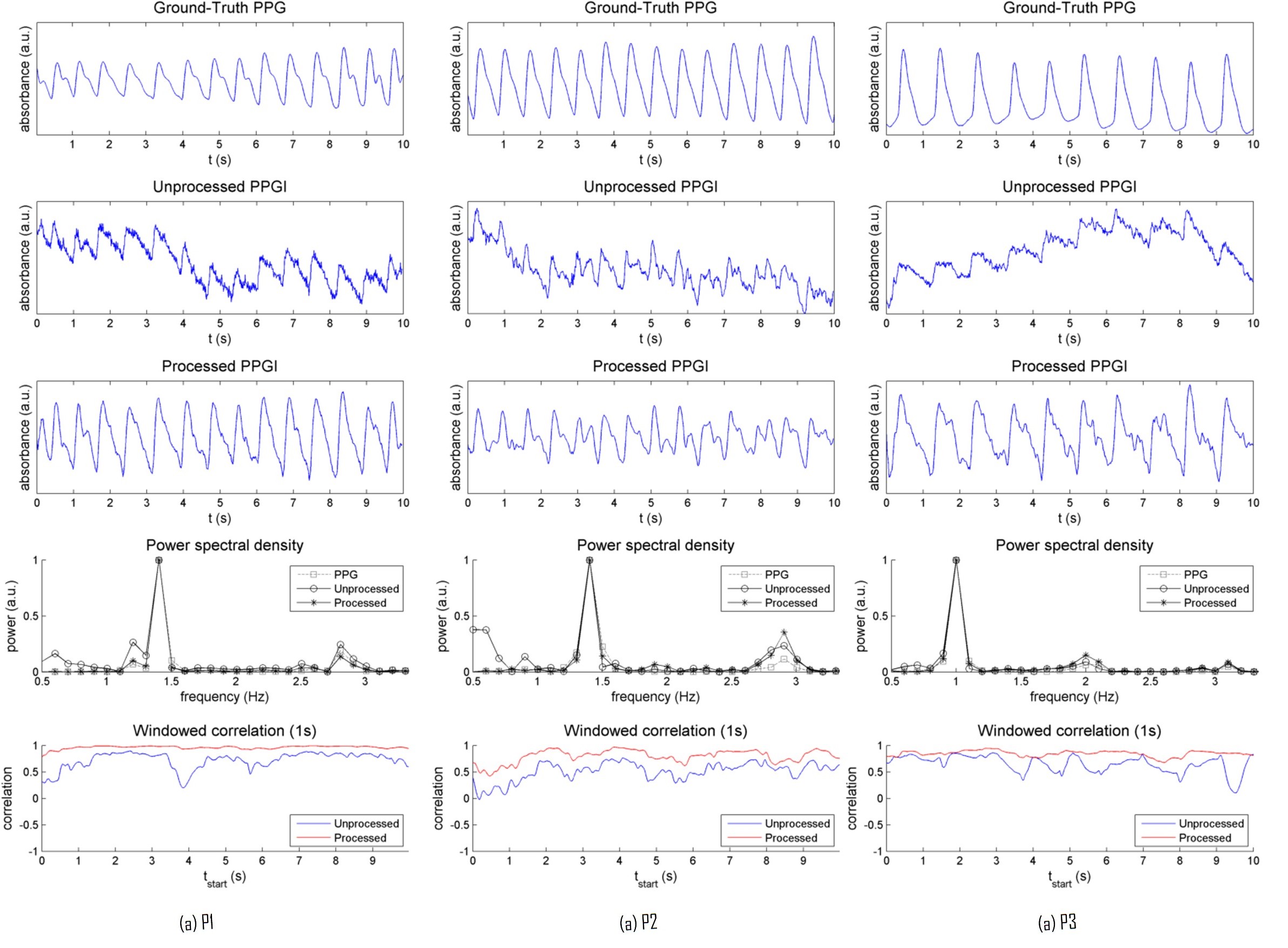}
\caption{Summary of results for Experiment 1 (short-distance measurement) across three participants (P1, P2, P3). The unprocessed acquired signal (row 2) was processed using ambient correction and signal processing, yielding a PPGI signal (row 3), which exhibited a higher correlation to the ground-truth contact PPG signal (row 1). For each participant, the PPGI power spectral density (row 4) closely matched the ground-truth PPG power spectral density, and the heart rate was easily distinguished by the maximum peak. The windowed correlation (row 5) of the processed PPGI signal to the PPG signal (red) showed a unanimous improvement over the unprocessed signal (blue), exemplifying the local temporal similarity across the entire signal.}
\label{fig:results_e1_sigs}
\end{sidewaysfigure}

The goal of Experiment 1 was to validate the base case short-distance setup required prior to long-distance measurement. Figure~\ref{fig:results_e1_sigs} shows the effect of processing the PPGI signal for three participants. The processed PPGI signal contained amplified pulsatility over the unprocessed (``raw'') PPGI. Pulsatile signals were extracted for all five participants (``P1'' through ``P5''), resulting in discernible systolic peaks of each blood pulse. The processed PPGI subdued much of the noise present in the raw PPGI signal. The high-frequency information not expected in the naturally smooth blood pulse waveforms was denoised, yielding smooth blood pulses. Furthermore, non-linear trends were corrected, yielding stable PPGI signals. These non-linear trends resulted in low correlation values between the raw PPGI and the ground-truth PPG (e.g., at 4~s for P1). The processed PPGI stabilized the signal, yielding larger correlation values at those time windows. The heart rate was discernible in both raw and processed PPGI as a peak frequency power coincident with the PPG peak. However, although all PPGI signals showed strong spectral similarity to the PPG signal in the PSD, the processed PPGI signal exhibited a stronger peak at the heart rate, thus subduing extraneous frequencies.


Table~\ref{tab:results1} summarizes the statistical results for this experiment. The correlation values across whole-signal comparisons were more than doubled for each participant from unprocessed ($\rho = 0.24 \pm 0.10$) to processed ($\rho = 0.72 \pm 0.19$) due to recovered stability and pulsatility. This indicates a large correction across the entire signal, demonstrating an overall improvement in signal fidelity. Statistical analysis using the heart beat instance correlation values showed a statistically significant increase in correlation for each participant ($p \in [<0.0001, 0.040]$). Thus, as a result of processing, the system was able to improve the PPGI signal such that there was a significant increase in pulsatility beat-to-beat, making inter-beat analysis more feasible.

\subsection*{Experiment 2: Long-Distance}
\label{sec:results_ld}

\begin{sidewaysfigure}
\centering
\includegraphics[width=0.95\textwidth]{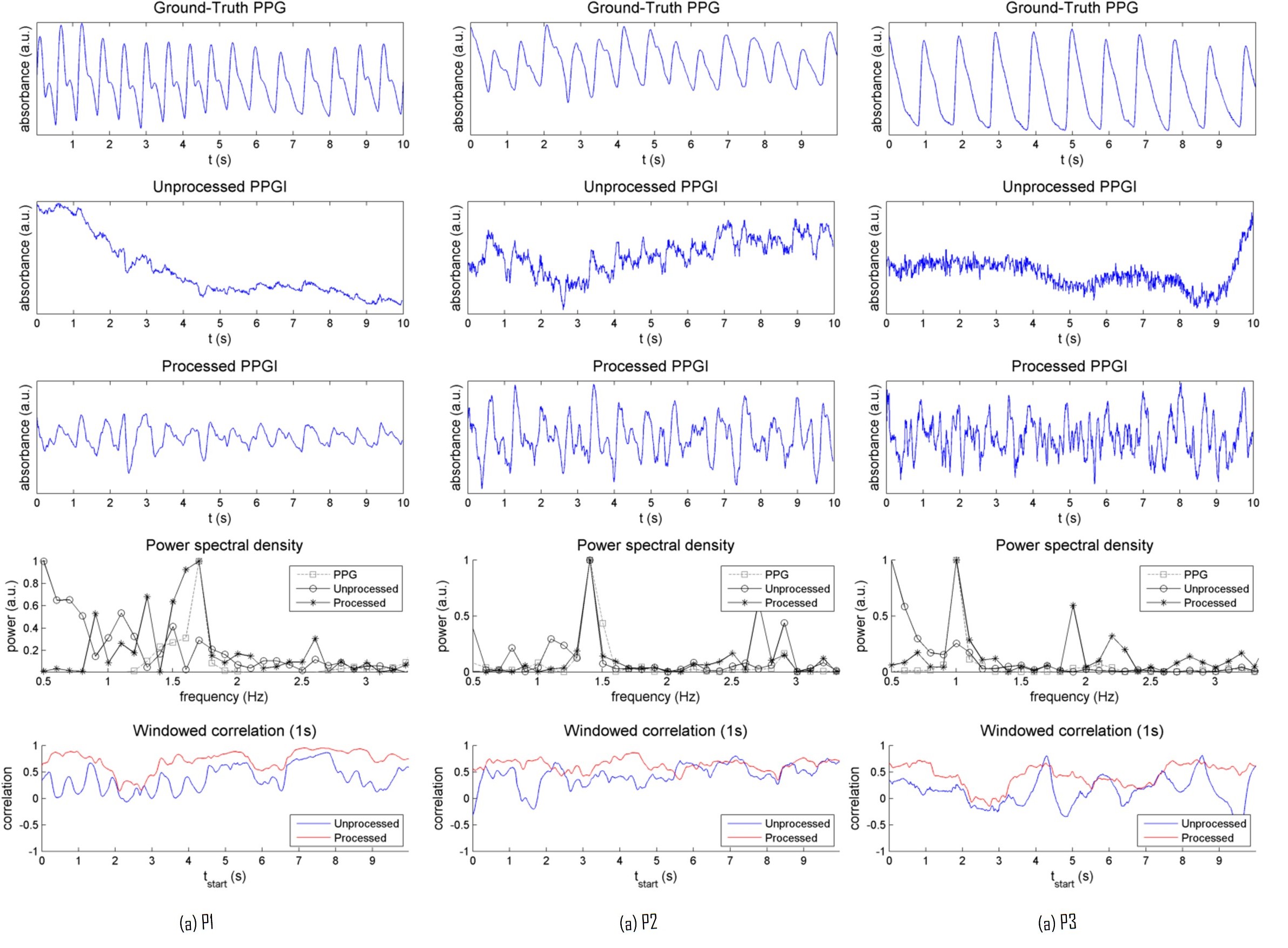}
\caption{Summary of results for Experiment 2 (long-distance measurement) across three participants (P1, P2, P3). The unprocessed acquired signal (row 2) was processed using ambient correction and signal processing, yielding a PPGI signal (row 3), which exhibited a higher correlation to the ground-truth contact PPG signal (row 1). The PPGI system was able to extract some pulsatile information from non-pulsing (P1) and noisy (P2, P3) signals. As a result, the heart rate was determined through the maximum frequency power peak (row 4), when such a peak may not have existed in the unprocessed signal (P1, P3). The windowed correlation (row 5) of the processed PPGI signal to the PPG signal (red) showed an overall improvement over the unprocessed signal (blue), exemplifying the local temporal similarity across the entire signal.}
\label{fig:results_e2_sigs}
\end{sidewaysfigure}

The goal of Experiment 2 was to validate long-distance PPGI measurement for long-distance monitoring. Figure~\ref{fig:results_e2_sigs} shows the effect of processing the PPGI signal for three participants. The raw PPGI signals were noisy, and pulsatility was not easily discernible in the signals of P1 and P3. This was emphasized by the lack of a clear peak in the PSD. Processing recovered the pulsatility for these participants, and yielded a less noisy and more stable signal. The processed PPGI signals yielded a higher degree of pulsatility, although the signals were noisier than the short-distance measurements (this is discussed later). This pulsatility was reflected in the PSDs, where there was a clear peak coincident with the heart rate for each participant. However, in cases such as P2 and P3, distinct blood pulses were not easily observed, making beat-to-beat timing analysis difficult at long-distances. The correlation plots showed a general increase across each participant with processing, demonstrating the increased pulsatility. However, there were some areas that were not improved (e.g., P3 at 2--3~s), largely due to unsuppressed noise likely due to amplified movement artifacts (discussed later). This high-frequency noise did not affect the PSD within the physiologically valid heart rate range.

Table~\ref{tab:results2} summarizes the statistical results for this experiment. As in Experiment~1, the correlation values for whole-signal results more than doubled from unprocessed ($\rho = 0.14 \pm 0.05$) to processed ($\rho = 0.55 \pm 0.12$), indicating recovered pulsatility and stability for an overall increase in signal fidelity. Statistical analysis using the heart beat instance correlation values showed a statistically significant increase in correlation for each participant except P3 ($p \in [<0.0001, 0.056]$). However, P3 exhibited substantial improvement upon processing ($p = 0.056$). This emphasizes the increased beat-to-beat recovery of pulsatile components. Overall, correlation values were lower than for short-distance monitoring, but significance (or near-significance) was achieved.

\section*{Discussion}
\label{sec:discussion}
Strong results were obtained for short-distance measurement results, producing visually pulsatile signals and accurately extracting heart rate. Statistical significance was achieved ($p \in [<0.0001, 0.040]$), exemplifying the recovery of pulsatile PPGI signals from a transmittance signal acquired in a non-contact manner. The processed PPGI signal yielded smooth pulsatile components, making temporal beat-to-beat analysis feasible. For example, heart rate variability can be used as a measurement for autonomic nervous system activity~\cite{berntson1997,stergiou2014}. The variability in correlation values across participants may be due to factors such as movement artifacts, and differences in scattering and absorptive media across tissues, such as melanin (skin tone) and adipose tissue~\cite{tuchin2007}. Gender differences were not analysed, since the participants were all male. Overall, the results validate short-distance cardiovascular monitoring using PPGI.

Long-distance monitoring yielded lower correlation values to ground-truth PPG than those for short-distance monitoring. Due to the diffuse nature of light traveling through highly scattering tissue~\cite{wang2012}, the sensor irradiance is reduced, yielding weaker signals at greater distances. Furthermore, the spatial resolution of a pixel increases with distance, and thus small movement artifacts were amplified. This is apparent in the results for participant P3 (see Figure~\ref{fig:results_e2_sigs}). The participant's fingers were shaky, yielding high-frequency noise which the system was unable to fully remove. During long-distance monitoring, statistical significance was achieved for heart beat instance correlation values in all but one participant (P3, $p=0.056$), indicating the recovery of a pulsatile PPGI signal.

Although pulsatility was extracted in some but not all cases, the fundamental heart rate was identifiable across all participants. This indicates that beat-to-beat timing analysis becomes difficult in the time domain, but heart rate is identifiable during long-distance PPGI monitoring. These results indicate the feasibility of long-distance heart rate monitoring, but further work must be done to demonstrate beat-to-beat temporal analysis in long-distance monitoring. Such an imaging system could benefit settings in which the individual exhibits low motion, such as cardiovascular monitoring during surgical procedures or sleep studies. In both settings, having as little contact on the individual's body to limit the amount of obstruction is preferable. Such an imaging system could provide physiologically relevant information in a non-contact manner. Additionally, the imaging system could monitor multiple individuals within the camera's field of view by analysing each individual's signal independently. The cost per individual decreases with a greater number of individuals, since the setup requires a single camera, and one LED per individual.

\section*{Methods}
\label{sec:methods}
To investigate the feasibility of long-distance PPGI monitoring, a non-contact PPGI system is proposed whose design integrates hardware (active illumination, a camera, and synchronization electronics) and software (signal processing, timing signals). Figure~\ref{fig:system_diagram} depicts a graphical overview of the proposed PPGI system. A participant's fingers were placed between a high-power LED (Philips LUXEON Rebel, 655~nm, 580~mW) and a camera (PointGrey Flea3, 100~fps), enabling long-distance monitoring. The camera frames and LED illumination patterns are synchronized by a microcontroller (MCU) and custom electronics. Temporally coded illumination (TCI) is proposed for controlling this illumination pattern. Figure~\ref{fig:system_flow} shows the stages of processing to yield the final PPGI signal. The following subsections provide detailed description of the processing steps.

\subsection*{Light Transport Model}
A temporal extension of the Beer-Lambert law (BLL) was used. The BLL shows that light is exponentially attenuates as it passes through a homogeneous light-absorbing medium:
\begin{equation}
  T=\frac{I}{I_0}=e^{-\epsilon l c}
\end{equation}
where $T$ is transmittance, $I_0$ is incident light, $I$ is transmitted light, $\epsilon$ is the tabulated molar extinction coefficient for the light-absorbing medium (e.g., oxyhemoglobin), $l$ is the photon path length, and $c$ is the concentration of the medium. Noting that path length for oxyhemoglobin is the dominant temporally changing parameter in heart beat analysis~\cite{shelley2007}, the temporal transmittance can be written as a function of time $t$:
\begin{equation}
  T(t)=\frac{I(t)}{I_0} = e^{-\epsilon \cdot l(t) \cdot c}
\end{equation}
The standard measurement signal for PPG analysis is absorbance, which is related to transmittance as $A=-\log(T)$.
Then, expressing this as temporal absorbance measurement yields:
\begin{equation}
  A(t) = -\log\left(I(t)\right)
  \label{eq:A}
\end{equation}

This model assumes constant ambient conditions. However, $I(t)$ may be affected by temporal changes in ambient illumination, potentially corrupting the subtle blood pulse signal. Thus, in the next section, an ambient correction method is proposed.

\subsection*{Temporally Coded Illumination (TCI)}
A TCI process is proposed to remove the effect of ambient light changes in the data, thus ensuring that the reflectance signal is due solely to the controlled active illumination (e.g., LED). A dual-mode temporal code was defined by which frames contain irradiance due to either ambient illumination or a combination of ambient and active illumination. By coding the illumination synchronously with frame captures, ambient measurements can be performed according to the temporal code, and subtracted from the actively illuminated frames. For example, a temporal code of 3:1 defines a TCI sequence that measures one ambient frame after every three actively illuminated frames. Figure~\ref{fig:system_diagram} shows a graphical example of 3:1 TCI. 

Mathematically, given an intensity image $I(t)$ at time $t$, according to the TCI sequence, $I(t)$ can either be:
\begin{equation}
I(t) = \begin{cases}
  I_{amb}(t), &\text{LED off}\\
  I_{amb}(t)+I_{act}(t), &\text{LED on}
\end{cases}
\end{equation}
where $I_{amb}(t)$ and $I_{act}(t)$ are the irradiance due to ambient illumination and active illumination, respectively. Then, given a recently acquired ambient frame $I_{amb}(t-\Delta t)$, by modeling light reflectance as a linear combination of ambient and active lighting, the ambient correction can be applied to the current frame:
\begin{equation}
  I^*(t) = I(t) - I_{amb}(t-\Delta t)
  \label{eq:istar}
\end{equation}
With a fast frame rate, $\Delta t$ becomes negligible, yielding illumination due solely to controlled active illumination:
\begin{equation}
  \lim_{\Delta t \rightarrow 0} I^*(t) = I_{act}(t)
\end{equation}
Since the system camera operates at a frame rate of 100~fps, $\Delta t$ approximately satisfies this condition ($\Delta t=10$~ms). Substituting (\ref{eq:istar}) into (\ref{eq:A}) yields the ambient-corrected absorbance measurement:
\begin{equation}
  A(t) = -\log\left(I^*(t)\right)
\end{equation}
Once the ambient frame $I_{amb}(t)$ is captured, the previously acquired active frame $I(t-\Delta t)$ was used in its place to ensure constant frame rate for frequency analysis.

TCI was implemented at the hardware level, and requires control of an active light source. Thus an active illumination system was chosen. Camera and illumination synchronization was performed using custom circuitry and microcontroller code. Figure~\ref{fig:results_imgs} shows the results of ambient correction via TCI on a single frame. Media~1 shows ambient correction across a video with varying lighting.

\subsection*{Signal Processing}
Capturing data using the given light transport model and TCI yields an ambient-corrected PPGI signal. However, this signal may still be affected by various sources of noise. This section proposes a signal processing pipeline for extracting a stable pulsatile PPGI signal. First, a denoising step is proposed, which reduces the noise based on camera and process characteristics. Second, a detrending step is proposed, which removes trends in intensity due to movement.

\subsubsection*{Denoising}
The goal of signal denoising is to recover the smooth blood pulse waveform from a noisy signal. Various sources of noise may corrupt the signal, such as light shot noise, the inherently stochastic nature of light-tissue interaction, due to measurement noise due to camera read noise. Kalman filtering~\cite{kalman1960} models observed measurements as a signal containing two general types of noise: process noise and measurement noise. This models the system well, considering the light-tissue interaction as process noise and sensor noise as measurement noise. It is a recursive algorithm that can operate in real-time, as denoising at time $t$ depends only on the state and uncertainty matrix at time $t-1$. Since blood pulse waveforms are inherently smooth in nature, the system state is defined using Newton's laws of motion using the raw camera intensity values:
\begin{equation}
  \vec{x}_k = \begin{bmatrix}
    I_k \\
    \dot{I}_k
  \end{bmatrix} = \begin{bmatrix}
    1 & \Delta t \\
    0 & 1
  \end{bmatrix} \vec{x}_{k-1} + \begin{bmatrix}
    \frac{\Delta t^2}{2} \\
    \Delta t
  \end{bmatrix} \ddot{I}_k
\end{equation}
where, at time point $k$, $I_k$ is pixel intensity, $\dot{I}_k$ is the rate of pixel variation, and $\ddot{I}_k$ is acceleration of pixel intensity. The measurement noise was assessed by imaging a dark field and quantifying the standard deviation of pixel intensities $\sigma_m$ across the sensor. During the experiments, $\sigma_m=1.02$ for 8~bit pixel values and 8~ms integration time. Denoising was performed on the raw camera signal $I(t)$ rather than the computed absorbance signal $A(t)$ since $\sigma_m$ was learned using the untransformed raw camera intensity values.

Process noise is more difficult to model due to the complex and stochastic interaction of light with tissue. Thus, a grid search optimization was performed to assess the optimal process noise variance value $\sigma_p$ using 8~ms exposure time. Using a fixed measurement noise model and a training data set, a logarithmic grid search was performed over various values for $\sigma_p$. The parameter value that yielded the highest correlation to the ground-truth PPG signal was chosen, resulting in $\sigma_p=1.84\times10^{-2}$ for 8~bit pixels. This denoising process was then performed on the acquired signal prior to any further processing. Then, the denoised intensity signal was extracted using an observation matrix:
\begin{equation}
  \hat{I}_k = H\vec{x}_k
\end{equation}
where $H=[1~0]^T$. Plugging this into (\ref{eq:A}) yields the denoised absorbance signal:
\begin{equation}
  A(t) = -\log(\hat{I}(t))
\end{equation}

\subsubsection*{Detrending}
\begin{figure}[H]
\centering
\includegraphics[width=0.95\textwidth]{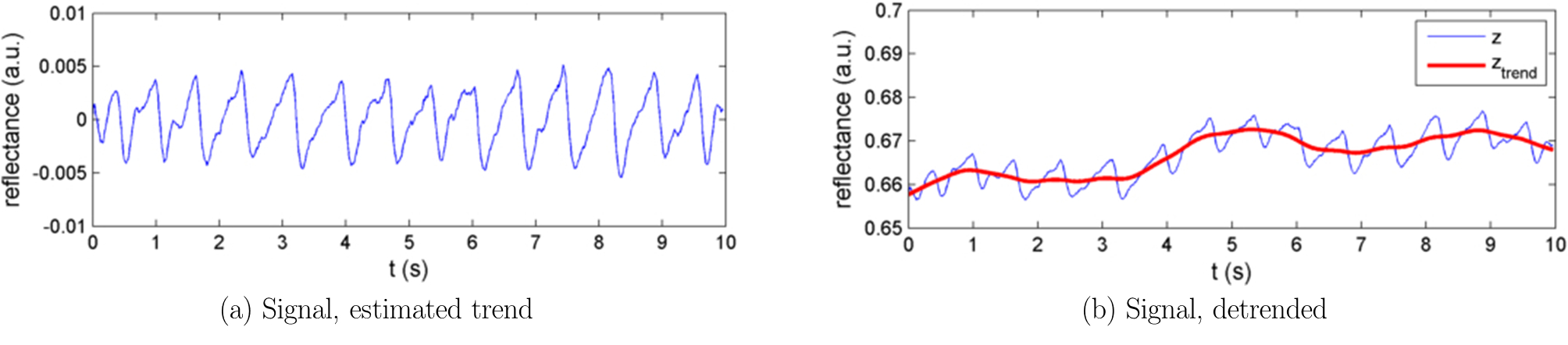}
\caption{Example of the detrending procedure. A slow trend is estimated in the original signal, and subsequently removed to obtain a signal with a stable mean. This process removes variability in illumination due to movement and subtle changes in incident light.}
\label{fig:detrend}
\end{figure}

Fluctuations in lighting and measurement or process noise should be corrected using the denoising algorithm. However, these corrections assume constant incident illumination, which is an invalid assumption during movement. The goal of this detrending step is to remove slow oscillations in the signal, yielding a PPGI signal with constant offset. It was assumed that movements are relatively smooth over time. A detrending algorithm was used on the denoised absorbance signal $A(t)$, in which the model assumes a smoothness prior~\cite{tarvainen2002}. In particular, the observed signal $A(t)$ is modeled as the linear combination of the ``true'' absorbance signal $A_{true}(t)$ and a temporal trend $A_{trend}(t)$.
\begin{equation}
  A(t)=A_{true}(t)+A_{trend}(t)
\end{equation}
Given that $A(t)$ is measured, $A_{true}(t)$ can be solved by estimating $A_{trend}(t)$ assuming a linear model, subtracting it from $A(t)$, and solving this using regularized least squares, which provides the following estimate of the true detrended signal:
\begin{equation}
  \hat{A}_{true}=\left(I-(I+\lambda^2 D_2^T D_2)^{-1}\right)A
\end{equation}
where $I$ is the identity matrix, $\lambda$ is a relative weighting term, and $D_2$ is the discrete approximation matrix of the second derivative. Figure~\ref{fig:detrend} shows the recovery of a stable signal from a signal corrupted by movement using this detrending method.

\section*{Acknowledgements}
\label{sec:ack}
The study was funded by the Natural Sciences and Engineering Research Council (NSERC) of Canada and, in part, by the Canada Research Chairs program.

\section*{Author Contributions}
R.A., C.S., F.K., A.W. designed the system. R.A. built the system. R.A., C.S., A.W. designed the algorithms. R.A. developed the data acquisition protocol. All authors were involved in data collection. R.A., K.P. designed the data analysis protocol. R.A., B.L. conducted data analysis. All authors reviewed the manuscript.

\section*{Additional Information}
\textbf{Competing financial interest:} All authors in this study have no competing financial interests.

%
%

\bibliographystyle{spiebib}
\bibliography{bib}

\begin{thebibliography}{10}

\bibitem{hertzman1937}
Hertzman, A.~B., ``Photoelectric plethysmography of the fingers and toes in
  man,'' {\em Experimental Biology and Medicine}~{\bf 37}(3),  529--534 (1937).

\bibitem{allen2007}
Allen, J., ``Photoplethysmography and its application in clinical physiological
  measurement,'' {\em Physiological Measurement}~{\bf 28}(3),  R1--R39 (2007).

\bibitem{shelley2007}
Shelley, K.~H., ``Photoplethysmography: Beyond the calculation of arterial
  oxygen saturation and heart rate,'' {\em Anesthesia \& Analgesia}~{\bf
  105}(6),  S31--S36 (2007).

\bibitem{levy2008}
Levy, B.~I., Schiffrin, E.~L., Mourad, J.-J., Agostini, D., Vicaut, E., Safar,
  M.~E., and Struijker-Boudier, H.~A., ``Impaired tissue perfusion a pathology
  common to hypertension, obesity, and diabetes mellitus,'' {\em
  Circulation}~{\bf 118}(9),  968--976 (2008).

\bibitem{wieringa2005}
Wieringa, F., Mastik, F., and Van~der Steen, A., ``Contactless multiple
  wavelength photoplethysmographic imaging: a first step toward ``{SpO$_2$}
  camera'' technology,'' {\em Annals of Biomedical Engineering}~{\bf 33}(8),
  1034--1041 (2005).

\bibitem{humphreys2007}
Humphreys, K., Ward, T., and Markham, C., ``Noncontact simultaneous dual
  wavelength photoplethysmography: a further step toward noncontact pulse
  oximetry,'' {\em Review of Scientific Instruments}~{\bf 78}(4),  044304
  (2007).

\bibitem{zheng2008}
Zheng, J., Hu, S., Azorin-Peris, V., Echiadis, A., Chouliaras, V., and Summers,
  R., ``Remote simultaneous dual wavelength imaging photoplethysmography: a
  further step towards 3-d mapping of skin blood microcirculation,'' in {\em
  Proc. SPIE}{\nolinebreak\hspace{0.1em}},   {\bf 6850},  68500S (2008).

\bibitem{cennini2010}
Cennini, G., Arguel, J., Ak{\c{s}}it, K., and van Leest, A., ``Heart rate
  monitoring via remote photoplethysmography with motion artifacts reduction,''
  {\em Optics Express}~{\bf 18}(5),  4867--4875 (2010).

\bibitem{kamshilin2011}
Kamshilin, A.~A., Miridonov, S., Teplov, V., Saarenheimo, R., and Nippolainen,
  E., ``Photoplethysmographic imaging of high spatial resolution,'' {\em
  Biomedical Optics Express}~{\bf 2}(4),  996--1006 (2011).

\bibitem{sun2011}
Sun, Y., Hu, S., Azorin-Peris, V., Greenwald, S., Chambers, J., and Zhu, Y.,
  ``Motion-compensated noncontact imaging photoplethysmography to monitor
  cardiorespiratory status during exercise,'' {\em Journal of Biomedical
  Optics}~{\bf 16}(7),  077010--1--077010--9 (2011).

\bibitem{verkruysse2008}
Verkruysse, W., Svaasand, L.~O., and Nelson, J.~S., ``Remote plethysmographic
  imaging using ambient light,'' {\em Optics Express}~{\bf 16}(26),
  21434--21445 (2008).

\bibitem{poh2010}
Poh, M.-Z., McDuff, D.~J., and Picard, R.~W., ``Non-contact, automated cardiac
  pulse measurements using video imaging and blind source separation,'' {\em
  Optics Express}~{\bf 18}(10),  10762--10774 (2010).

\bibitem{sun2012}
Sun, Y., Papin, C., Azorin-Peris, V., Kalawsky, R., Greenwald, S., and Hu, S.,
  ``Use of ambient light in remote photoplethysmographic systems: comparison
  between a high-performance camera and a low-cost webcam,'' {\em Journal of
  Biomedical Optics}~{\bf 17}(3),  037005--1--037005--10 (2012).

\bibitem{kong2013}
Kong, L., Zhao, Y., Dong, L., Jian, Y., Jin, X., Li, B., Feng, Y., Liu, M.,
  Liu, X., and Wu, H., ``Non-contact detection of oxygen saturation based on
  visible light imaging device using ambient light,'' {\em Optics Express}~{\bf
  21}(15),  17464--17471 (2013).

\bibitem{xu2014}
Xu, S., Sun, L., and Rohde, G.~K., ``Robust efficient estimation of heart rate
  pulse from video,'' {\em Biomedical Optics Express}~{\bf 5}(4),  1124--1135
  (2014).

\bibitem{anderson1981}
Anderson, R. and Parrish, J.~A., ``The optics of human skin.,'' {\em Journal of
  Investigative Dermatology}~{\bf 77}(1) (1981).

\bibitem{gibson2005}
Gibson, A., Hebden, J., and Arridge, S.~R., ``Recent advances in diffuse
  optical imaging,'' {\em Physics in Medicine and Biology}~{\bf 50}(4),
  R1--R43 (2005).

\bibitem{amelard2015}
Amelard, R., Scharfenberger, C., Wong, A., and Clausi, D.~A.,
  ``Illumination-compensated non-contact imaging photoplethysmography via
  dual-mode temporally-coded illumination,'' in {\em Proc.
  SPIE}{\nolinebreak\hspace{0.1em}},   {\bf 9316} (2015).

\bibitem{easypulse}
R-B, ``{Easy Pulse} sensor (version 1.1) overview (part 1).''
  http://embedded-lab.com/blog/?p=7336 (2013).
\newblock (12 Dec 2014).

\bibitem{berntson1997}
Berntson, G.~G., Bigger, J.~T., Eckberg, D.~L., Grossman, P., Kaufmann, P.~G.,
  Malik, M., Nagaraja, H.~N., Porges, S.~W., Saul, J.~P., Stone, P.~H., and
  van~der Molen, M.~W., ``Heart rate variability: origins, methods, and
  interpretive caveats,'' {\em Psychophysiology} (34),  623--648 (1997).

\bibitem{stergiou2014}
Stergiou, S. and Balakrishnan, R., ``Streaming updates for heart rate
  variability algorithms,'' {\em IEEE Transactions on Biomedical Engineering}
  (61),  1931--1937 (2014).

\bibitem{tuchin2007}
Tuchin, V.,  [{\em Tissue optics: light scattering methods and instruments for
  medical diagnosis}{\nolinebreak\hspace{0.1em}]}, SPIE Press Bellingham, 2~ed.
  (2007).

\bibitem{wang2012}
Wang, L.~V. and Wu, H.-i.,  [{\em Biomedical optics: principles and
  imaging}{\nolinebreak\hspace{0.1em}]}, John Wiley \& Sons (2012).

\bibitem{kalman1960}
Kalman, R.~E., ``A new approach to linear filtering and prediction problems,''
  {\em Journal of Fluids Engineering}~{\bf 82}(1),  35--45 (1960).

\bibitem{tarvainen2002}
Tarvainen, M.~P., Ranta-aho, P.~O., and Karjalainen, P.~A., ``An advanced
  detrending method with application to {HRV} analysis,'' {\em IEEE
  Transactions on Biomedical Engineering}~{\bf 49}(2),  172--175 (2002).

\end{thebibliography}

\newpage

\clearpage  

\newpage

\begin{table}[H]
  \caption{Experiment 1 statistical results for short-distance measurements. Pearson's linear correlation coefficient ($\rho$) was computed between the PPGI and PPG signals for the whole 10~s signal for global accuracy, and for each heart beat instance (i.e., diastole-to-diastole) for local accuracy. A $p$-value was computed using a paired-sample $t$-test to test the hypothesis that there is a zero-mean difference between unprocessed and processed correlation. At the whole-signal level, the proposed system processing yielded greater than doubly improved accuracy over the unprocessed signal. At the local level, statistically significant results ($\alpha=0.05$) were achieved for each participant, indicating significant improvement with processing.}
  \centering
  
  \begin{tabular}{|l|c|c|c|c|c|}
    \hline
     & P1 & P2 & P3 & P4 & P5 \\
    \hline
    \hline
    Unprocessed, whole & 0.35 & 0.31 & 0.26 & 0.14 & 0.12 \\
    Processed, whole & 0.94 & 0.79 & 0.82 & 0.49 & 0.56 \\
    \hline
    \hline
    Unprocessed, blocked$^*$ & 0.72 $\pm$ 0.16 & 0.51 $\pm$ 0.17 & 0.65 $\pm$ 0.16 & 0.22 $\pm$ 0.32 & 0.28 $\pm$ 0.36 \\
    Processed, blocked$^*$ & 0.96 $\pm$ 0.03 & 0.79 $\pm$ 0.12 & 0.85 $\pm$ 0.05 & 0.51 $\pm$ 0.21 & 0.59 $\pm$ 0.32 \\
    $p$-value & $<0.0001~(n=13)$ & $<0.0001~(n=14)$ & $<0.0001~(n=9)$ & $0.040~(n=10)$ & $0.020~(n=13)$ \\
    \hline
    \multicolumn{5}{l}{$^* \mu \pm \sigma$}
  \end{tabular}
  \label{tab:results1}
\end{table}

\begin{table}[H]
  \caption{Experiment 2 statistical results for long-distance measurements. Pearson's linear correlation coefficient ($\rho$) was computed between the PPGI and PPG signals for the whole 10~s signal for global accuracy, and for each heart beat instance (i.e., diastole-to-diastole) for local accuracy. A $p$-value was computed using a paired-sample $t$-test to test the hypothesis that there is a zero-mean difference between unprocessed and processed correlation. At the whole-signal level, the proposed system processing yielded greater than doubly improved accuracy over the unprocessed signal. At the local level, statistically significant results ($\alpha=0.05$) were achieved for each participant except P3 ($\rho=0.056$), indicating substantial improvement with processing.}
  \centering
  
  \begin{tabular}{|l|c|c|c|c|c|}
    \hline
     & P1 & P2 & P3 & P4 & P5 \\
    \hline
    \hline
    Unprocessed, whole    & 0.18 & 0.21 & 0.14 & 0.09 & 0.10 \\
    Process, whole      & 0.62 & 0.61 & 0.45 & 0.66 & 0.39 \\
    \hline
    \hline
    Unprocessed, blocked$^*$ & 0.38 $\pm$ 0.24 & 0.44 $\pm$ 0.21 & 0.13 $\pm$ 0.28 & 0.32 $\pm$ 0.20 & 0.12 $\pm$ 0.29 \\
    Processed, blocked$^*$   & 0.72 $\pm$ 0.19 & 0.63 $\pm$ 0.11 & 0.43 $\pm$ 0.22 & 0.70 $\pm$ 0.14 & 0.33 $\pm$ 0.36 \\
    $p$-value & $<0.0001~(n=15)$ & $0.0088~(n=13)$ & $0.056~(n=9)$ & $0.00040~(n=10)$ & $0.0079~(n=11)$ \\
    \hline
    \multicolumn{5}{l}{$^* \mu \pm \sigma$}

  \end{tabular}
  \label{tab:results2}
\end{table}

\end{document}